# A Practical Analysis of Rust's Concurrency Story

Aditya Saligrama
MIT PRIMES
saligrama@csail.mit.edu

Andrew Shen
MIT PRIMES
shenandrew95@gmail.com

Jon Gjengset
Parallel and Distributed
Operating Systems
MIT
jon@thesquareplanet.com

## Abstract

Correct concurrent programs are difficult to write; when multiple threads mutate shared data, they may lose writes, corrupt data, or produce erratic program behavior. While many of the data-race issues with concurrency can be avoided by the placing of locks throughout the code, these often serialize program execution, and can significantly slow down performance-critical applications. Programmers also make mistakes, and often forget locks in less-executed code paths, which leads to programs that misbehave only in rare situations.

Rust is a recent programming language from Mozilla that attempts to solve these intertwined issues by detecting data-races at compile time. Rust's type system encodes a data-structure's ability to be shared between threads in the type system, which in turn allows the compiler to reject programs where threads directly mutate shared state without locks or other protection mechanisms. In this work, we examine how this aspect of Rust's type system impacts the development and refinement of a concurrent data structure, as well as its ability to adapt to situations when correctness is guaranteed by lower-level invariants (e.g., in lock-free algorithms) that are not directly expressible in the type system itself. We detail the implementation of a concurrent lock-free hashmap in order to describe these traits of the Rust language. Our code is publicly available at <https://github.com/saligrama/concache> and is one of the fastest concurrent hashmaps for the Rust language, which leads to mitigating bottlenecks in concurrent programs.

# Approach

## Why concurrency?

Writing correct concurrent code is difficult. Once you cannot foresee exactly the order in which operations happen, and how operations get interleaved, it is hard to fully grasp all the ways in which your program may be wrong. Concurrent data structures in particular fall into this category, as there are often subtle invariants that must be maintained for correct behavior, and it's easy to miss corner-cases that produce incorrect program behavior.

The primary cause of headache in concurrent programs is *data races*. A data race occurs when two threads attempt to read and write the same memory location at the same time. This can lead to strange program behavior: a read operation that executes concurrently with a write can return the value from either before *or after* the write, and you don't know which. Similarly, if two writes occur concurrently, one of the two writes will be lost, but which one will be lost is also unknown.

Programmers usually avoid data races by inserting *locks*. A lock serializes the execution of threads that try to modify the same data, and ensures that they do not step on each other's toes. However, if the programmer is not careful, they may forget to <u>always</u> take the lock when accessing protected data, in which case the protection is ineffective. Because locks serialize access, they also present a scalability bottleneck: only one thread can access the data at a time, even if they are accessing different parts of it. *Fine-grained locking* helps remedy this by using different locks for different segments of the data, but often-accessed segments may still see significant slow-downs.

*Lock-free data structures* are data structures that have been designed specifically for concurrent access. Operations on such data structures do not generally need to take locks, and can proceed concurrently even when there are many readers and many writers. But writing correct lock-free algorithms is challenging, as every possible operation interleaving must be considered and designed for. One particular challenge that arises in lock-free data structures is freeing resources that have been deleted. Since any number of threads may also be reading deleted data as it is getting deleted, developers must take care to maintain invariants that allow them to detect when it is safe to deallocate removed data.

## Why Rust?

One of Rust's slogans is "fearless concurrency"; through the "borrow checker", a component of the Rust compiler, Rust attempts to make it easier to write *correct* concurrent code, and to eliminate data race bugs entirely. Specifically, Rust uses its type system to force programmers to ensure that concurrent mutations of their data is done in a safe manner. We discuss the exact mechanisms used to achieve this later, but in general, when the compiler cannot verify that the code meets constraints that are deemed necessary for safety, that code does not compile.

While Rust's type-system approach to eliminating data races works well for the majority of programs, developers who wish to implement sophisticated lock-free algorithms or data-structures may find the type system too conservative. For example, an algorithm may guarantee that memory accesses do not race by maintaining complicated invariants that the compiler cannot check. To allow developers to implement such advanced algorithms, Rust provides an escape hatch through *unsafe* code. Code that is marked as unsafe is allowed to alias and typecast pointers (though it must be valid Rust in every other way), which is sufficient to implement any concurrent algorithm.

In this work, we implement a hashmap, a very common data structure, and add increasingly sophisticated concurrency support to it. We then analyze how well the Rust type system protects us from errors, and where we need to use the unsafe escape hatch. We also investigate the ramifications of using unsafe, and the degree to which it impacts the other parts of the data structure implementation.

## Why hashmaps?

Hashmaps are ubiquitous, and can be found in most applications in one form or another. Since they often carry information that is accessed by many threads, they are also commonly used concurrently under the protection of a lock. Because of this, these maps can quickly become scalability bottlenecks as the application grows, and are prime candidates for optimization. As with most data structures, a hashmap can be modified to support increased concurrency in many ways, from the simple but inefficient "one big lock around the map" to more complicated fine-grained locking or lock-free designs. In this work, we examine multiple such implementations, and see how they interact with the Rust compiler and type system.

## Implementation strategy

To evaluate how Rust's type system helps when developing concurrent programs, and where its safety mechanisms become insufficient and have to be worked around, we implemented several concurrent hashmap designs of increasing sophistication and evaluated their performance. We started out using the hashmap provided by the standard library, wrapped in a reference-counted reader-writer lock, which, as expected, exhibited poor multi-core scalability. We then transitioned to a custom hashmap implementation with per-bucket locks, and finished with a mostly lock-free implementation that uses lock-free linked lists for each bucket and implements an epoch-based memory reclamation scheme. For the last design, we both wrote an implementation from scratch, and one that uses the concurrency library **crossbeam** to abstract away core concurrency primitives. Our implementation is available at

[https://github.com/saligrama/concache](https://github.com/saligrama/concache). We discuss our findings and the performance results in the following sections.

# A Rust concurrency primer

Rust uses several mechanisms to certify that concurrent programs are safe, many of which are not commonly found in other programming languages. We discuss each of them briefly here, but interested readers are encouraged to read the [Rust "book"](), the [`std::cell` documentation](), and the ["Nomicon" section on concurrency]().

In Rust, "safety" is generally defined as not being susceptible to undefined behavior, which occurs when compilers make certain assumptions that are not satisfied during execution. This behavior is difficult to debug and address, because it can vary between different systems and between executions of the same program. Rust attempts to point out potential bugs that could lead to undefined behavior, including out-of-bounds reads and writes, values that are used after being freed and null pointer dereferencing. In the context of concurrent code, we also consider thread safety and preventing data races, which occur when two or more threads access data such that one or more thread is a modifier. Below, we detail each of the aspects of Rust that contribute to checking a program for safety.

**Ownership:** In Rust, every variable is *owned* by some scope. When a scope ends, it is responsible for cleaning up any resources used by the variables that it owns. For example, when a pointer to heap-allocated memory leaves a scope, that automatically frees the allocated memory. Similarly, if a file is opened, the handle to that file will close the file when it goes out of scope. Each variable can only have one owner at a given time, but ownership can be passed to other scopes through function calls or returns. Variables that have gone out of scope cannot be accessed any more (e.g., no dangling pointers), and this is checked at compile-time.

**References and Borrowing**: Rust allows the owner of a variable to give out temporary *references* to a variable (this is called *borrowing* the variable). These are similar to pointers in other languages, but are also annotated with a *lifetime* which tracks how long the pointed-to value is in scope, and therefore how long the reference is valid for. The compiler checks that a reference is never used after its lifetime expires, which ensures that an object is never touched after it has been deallocated. There are two types of references in Rust: mutable (`&mut T`) and immutable (`&T`). Only a single mutable reference can exist at any given point in time, whereas any number of immutable references may exist simultaneously. This is also checked at compile-time. Values can only be modified through mutable references (even recursively, as opposed to C's `const`), which ensures that there can be no data races: since there cannot be both a mutable reference and immutable reference active at the same time, data cannot be modified while you are reading it.

**Send and Sync:** While the borrow checker (the part of the compiler that checks that all references are valid) guarantees that there are no data-races, additional mechanisms are needed to ensure that multithreaded programs behave correctly. Consider the case of a reference-counted variable. If the reference counting is not done using atomic instructions, it is not safe to have pointers to that value in multiple threads at the same time; the reference count could end up incorrect if both threads try to modify the count simultaneously. Rust also has some types that provide *interior mutability*: certain types allow you to modify a variable even if you only have an immutable reference to it. For example, the `Cell` type allows you to swap the value of a variable through a `&Cell`, which is safe as long as the `Cell` is only accessed from a single thread. To check these kinds of rules, Rust has two *marker traits*, which are a bit like compiler hints, called `Send` and `Sync`. A type that is `Send` can safely be sent to another thread (that is, its ownership can be passed across thread boundaries), whereas a type that is `Sync` can safely be accessed from another thread (that is, a reference to it can be passed across thread boundaries). A type whose members are all `Send` is itself `Send`, and the same applies to `Sync`. A non-atomic reference-counted variable (`Rc` in the standard library) is neither `Send` nor `Sync`, whereas a `Cell` is `Send` but not `Sync`. Rust requires code that is spawned on a new thread to be `Send`, which ensures that threads are not able to access shared data unless that data is contained in a structure that allows concurrent access.

**Unsafe code:** Rust also offers a fallback for programs whose safety cannot be expressed using ownership, references, `Send`, and `Sync`: the `unsafe` keyword. A code block that is marked as unsafe is allowed to create raw pointers — pointers without an associated lifetime — and cast them to different types, or back to regular references. This allows the developer to maintain multiple mutable pointers to the same data, and expose them as mutable references when their manually-checked invariants indicate that doing so is safe. This is necessary to implement, e.g., a lock, which exposes a mutable reference only when it has checked that there is no-one else holding the lock. It is important to note that unsafe is not simply a way to compile bad code; most of the type system still functions within unsafe blocks, and invalid programs where types do not match or regular borrow rules are violated are rejected. An important aspect of unsafe blocks in Rust is that they can also encapsulate unsafe behavior within a safe interface. Users of a library that contains unsafe code do not have to mark their own code as unsafe; the library authors effectively promise that their library provides a safe external interface. For example, while a lock uses unsafe code internally, no unsafe code is needed to use a lock.

# Observations and Results

## Performance

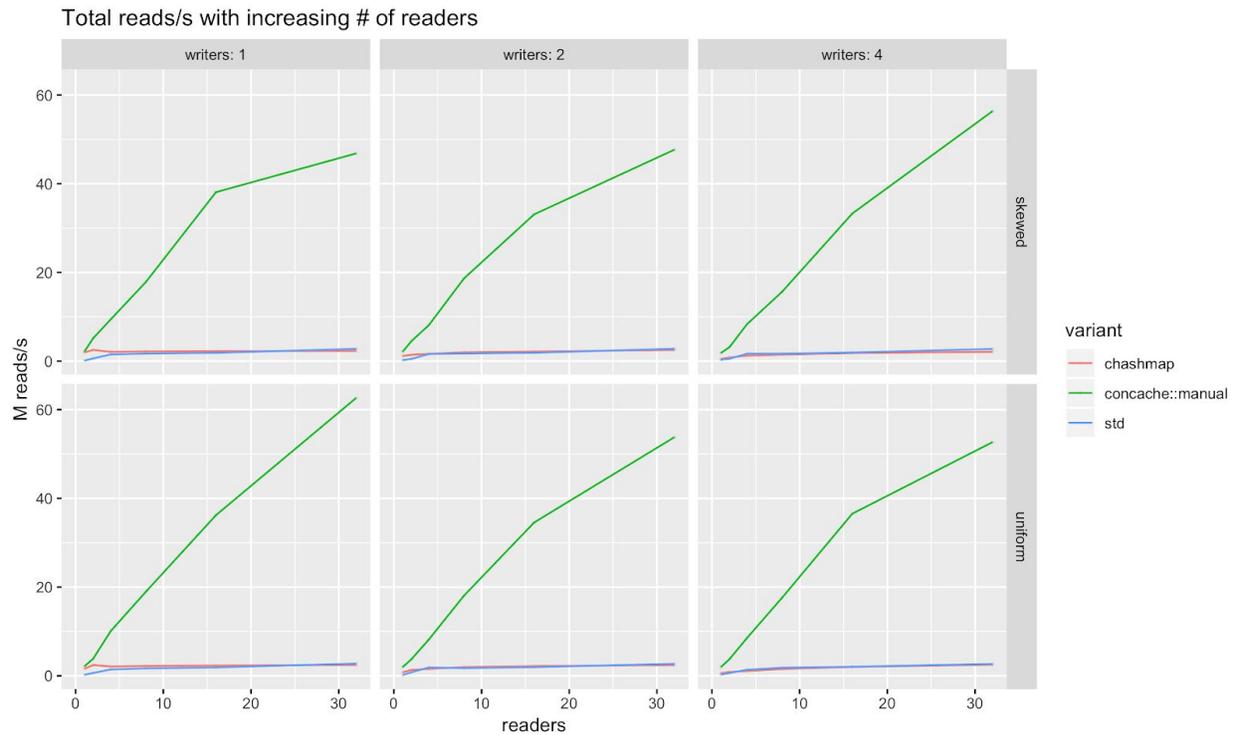

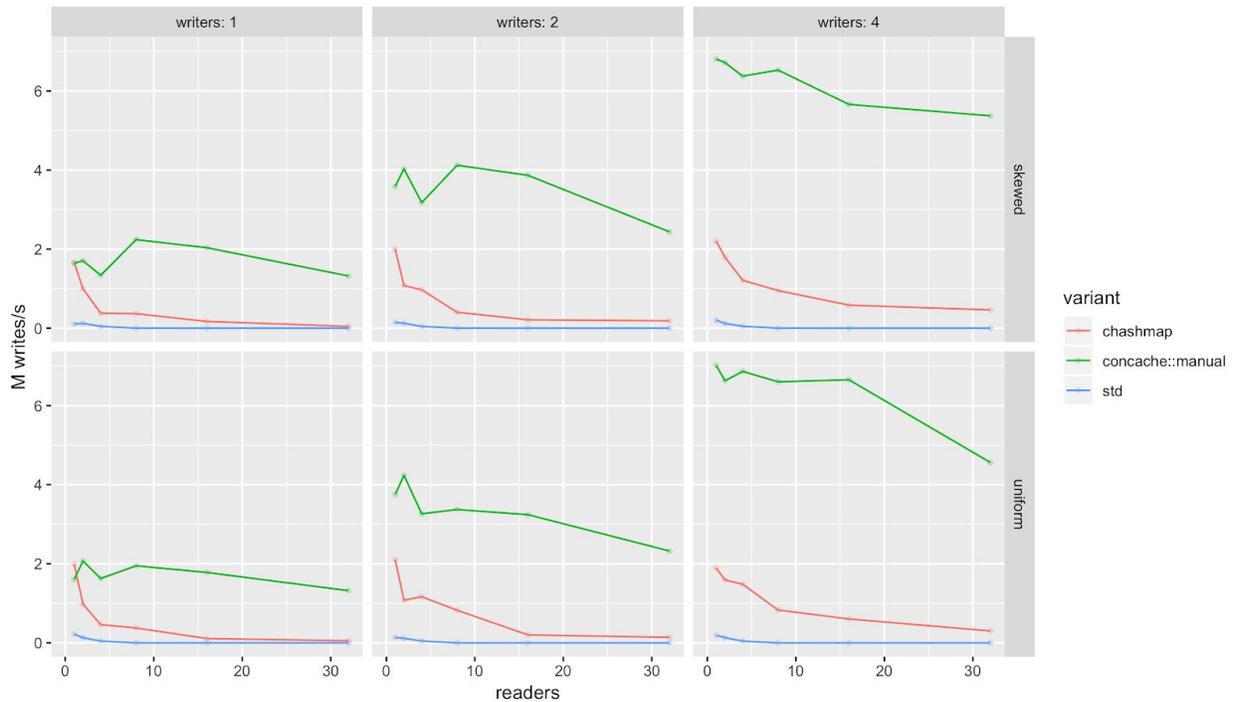

Total writes/s with increasing # of readers

## Analysis

Our results showed significant improvements over other naive implementations of concurrent hashmaps, namely chashmap and the standard hashmap wrapped in a lock. Concache performed exceptionally well with increasing numbers of readers and writers with a nearly linear increase in reads per second. In comparison, it had mediocre performance with respect to the number of writes per second as the number of writes per second decreased gradually as the number of readers and writers increased. This is likely due to the increase wait in accessing an element in the hashmap and is expected as a trade off against the high numbers of readers that concache is capable off. We are still unsure about the peculiar dip in the graph when measuring the number of writes per second, but we suspect that it is due to either the way concache uses the CPUs or spawns threads.

# Experience

In this section, we describe our experiences with using the Rust language to write concurrent hashmaps, detailing where the compiler helped and got in the way as well as other issues we found with the compiler and type system.

## Where Rust helps (Good)

In this subsection, we describe instances where the Rust model and type system helped us implement tricky parts of our code and in some cases prevented us from writing dangerous code which might produce runtime bugs or cause the program to behave unexpectedly.

**Explicit protection via locks:** In Rust, locks explicitly wrap around the type they are protecting. For example, a `Mutex<T>` protects the inner type `T` and forces threads to hold the lock to access that inner type. Conversely, in other languages such as Go, where the lock is separate from the inner type and threads are not forced to take the lock to access data.

**Lifetimes for epochs:** Crossbeam provides an epoch type that it uses to ensure that concurrent operations are executed in a safe order. In our Crossbeam implementation, we pinned epochs each time we performed a concurrent insert, delete or search operation on the hashmap. The Crossbeam epoch type harnessed the lifetime feature, forcing us to pin the epoch in the correct place so it outlived atomic values that were read.

**Auto-Free:** We have to worry less about managing memory as Rust will automatically free memory and call the appropriate deconstructor when a variable leaves scope. We found this to be very useful when writing *safe* code.

```
fn foo() {
    let n = 5usize;
    //n leaves scope here, the memory is automatically freed
}
```

**Ownership and Mutability:** Rust has very strict rules around ownership and mutability, and the compiler forces you to abide by those rules. This can act as a barrier early on, and make it harder to make your code compile initially, but also eliminates many bugs from the code once it does. Specifically, in safe code, Rust will ensure that you never use memory after freeing it, that you do not modify values while you are reading or modifying them elsewhere, and that you do not accidentally modify values that you have declared to be immutable. Once you have learned the rules that Rust imposes on you, they also start to feel somewhat natural, and you often write better code as a result.

**Similarity to C:** After completing the Rust implementation of the concurrent linked list, we noticed, with some minor exceptions (e.g the use of Option), that the Rust language was flexible enough to match the C-style pseudo code fairly well. This is particularly impressive because of the limitations that Rust enforces in order to maintain its safety features. It is also very useful as many concurrent algorithms are written in C-style pseudo code making it easier to port these algorithms into Rust.

**Pointers Rather than References:** When writing concurrent code, you often need to temporarily violate the Rust safety restrictions, or guarantee them through invariants that the compiler cannot check (e.g., multiple mutable pointers to the same data, or pointer manipulations). In these cases, Rust forces you to explicitly mark that code using the unsafe keyword. This forces the developer to realize that they are trying to do something that they need to think carefully through, but also makes it easier to narrow down where concurrency bugs might manifest later.

## Where Rust adds confusion/complexity (Bad)

In this subsection, we describe instances where we struggled against the compiler and the type system. The problems listed below generally caused our code to crash and were not immediately obvious to debug.

**Auto-Free:** Conversely, auto-free tends to be somewhat difficult when writing *unsafe* code as we must be careful to not accidentally drop a temporarily owned item (in our case, through Box::from_raw). Additionally, we must mediate the calling of the deconstructor and freeing of memory for memory that is not auto-freed (e.g when we call `Box::into_raw` on a Box<T>).

```
// incorrect -- node gets freed every time foo is called
fn foo(node: *mut Foo) -> usize {
  let x = unsafe { Box::from_raw(node) };
  // makes a lot of sense since inverse of Box::into_raw()
  // do something with x.next
  x.value
  // ERR: memory x points to gets freed since drop(Box<T>)
  // free(x)
}

// one fix: but inconvenient
fn foo(node: *mut Foo) -> usize {
  let x = unsafe { Box::from_raw(node) };
  // do something with x.next
  let v = x.value;
  mem::forget(x); // don't free the Box
  // no call to free(x) !
```

```
      v
}

// better: drop(x) does not free anything (x is a &T, not a Box<T>)
fn foo(node: *mut Foo) -> usize {
  let x = unsafe { &*node };
  // do something with x.next
  x.value
}
```

**Function Parameter Type:** When we implemented our linked list, we had to write a search function which returns a `Node`. However, deciding the exact type of the return can be tricky and Rust would allow any of three choices, if not more: `AtomicPtr<T>`, `*mut T`, `&mut T`. It is not immediately clear if any have an advantage over each other or even if they are at all different. In addition, combinations of these types are also possible.

**Pointer Manipulation:** In Rust, pointer arithmetic can be used to manipulate pointers for a variety of purposes such as encoding information into pointers. However, there is little to no safety in Rust regarding pointer manipulation. We are allowed to freely modify and change raw pointers and can dereference them in an unsafe block which will throw an illegal instruction if this memory is inaccessible. Rust does not help in determining whether a pointer derived from a specific piece of memory has been manipulated since its creation.

One suggestion might be to create a new pointer type that indicates when a pointer becomes manipulated making it possibly unsafe to dereference. To prevent accidental, we must manually declare the raw pointer as valid. In the example below, we call this new pointer type *addr and implement some of its basic features.

```
let x = Box::new([0; 8192]);
let ptr = Box::into_raw(x);
ptr.add_offset(200);
// require specific function for turning *addr -> *mut
// all dereferencing functions take *mut
let z = unsafe { &*ptr }; // ERR: ptr is *addr!
let z = unsafe { &*mem::declare_valid(ptr) };
// where mem::declare_valid = fn<T>(*addr T) -> *mut T {}
// get_marked_reference = fn(*mut T) -> *addr T { ptr |= 0x1 }
// get_unmarked_reference = fn(*addr T) -> *mut T {
//   mem::declare_valid(ptr ^ !0x1)
// }

let x = Box::new(8usize);
let ptr = Box::into_raw(x);
```

```
let ptr2 = (ptr as usize) as *mut usize;
let z = unsafe { Box::from_raw(ptr) };
drop(z);
let w = unsafe { Box::from_raw(ptr2) };
drop(w); // CRASH!
```

## Where Rust gets in the way (Ugly)

In this subsection, we describe instances where we had to work around the type system and the compiler in order to make our code compile. The points listed here are not necessarily problems with the Rust model and type system, but instead are ergonomic issues that make writing Rust code more painful than it needs to be.

**Prevalence of option and result:** Many Rust functions return a `Option<T>` or `Result<T>` so that the program does not use `None` values or so that execution does not stop when an error is thrown. The `unwrap()` function is a way to access the inner value without dealing with the `None` or `Err` cases; however the program will panic if a `None` or `Err` value is unwrapped. In order to safely handle `Option` or `Result`, we use a match statement as follows.

```
fn get_val() -> i64 {
    return match function_returning_result() {
        Ok(t) => t,
        Err(e) => {
            println!("Error occurred! {:?}", e);
            -1
        }
    };
}
```

We observed a tendency to overuse `unwrap()` while prototyping as it was easier to do so, shortening the length of the code. However, as we transitioned to finalizing our code and implementing better error handling, we found that it took some effort to remove these unwraps and replace them with match statements or the equivalent. While we recognize that robust error handling is difficult to implement, we would appreciate shorter and more efficient ways to transition from prototype to final code in terms of error handling.

**Unhelpful compile errors:** In several cases, the Rust compiler outputs certain compile errors that can be confusing to understand. A common example is error E0597 "Borrowed value does not live long enough." An example of such code can be found in the [Rust error index](#):

```
struct Foo<'a> {
    x: Option<&'a u32>,
}
```

```
let mut x = Foo { x: None };
let y = 0;
x.x = Some(&y); // error: `y` does not live long enough
```

Compiler output:

```
error[E0597]: `y` does not live long enough
 --> src/main.rs:8:17
  |
8 |     x.x = Some(&y);
  |                 ^ borrowed value does not live long enough
9 | }
  | - `y` dropped here while still borrowed
  |
  = note: values in a scope are dropped in the opposite order they are created
```

https://play.rust-lang.org/?gist=02990922a0b883882ed526c01fd87eb1&version=stable&mode=debug&edition=2015

This does not compile because values go out of scope in the opposite order that they are instantiated. Specifically, y goes out of scope before x does, but in this code example x depends on y. However, error E0597 often displays too little explanation for some developers, especially Rust beginners, to fix the problem. A more expanded and detailed error message here would be useful. The displayed error does not explicitly mention that x is part of the problem, and this understanding is necessary to fix the issue.

**Overuse of atomics:** In writing concurrent code, atomics are often needed in order to guarantee some level of consistency between concurrent operations. Rust's compiler often implies that one should use atomics through the E0597 "Borrowed value does not live long enough" error mentioned above. However, this conditions the developer to use atomics everywhere in concurrent code. For example, in our Crossbeam lock-free implementation, we initially used an AtomicUsize to store the number of buckets in the enclosing HashMap class due to this conditioning. However, because this value would only be mutated by a single thread during resizing, we eventually realized that this was unnecessary. We would ideally like to see a compiler feature, perhaps in `cargo lint`, that could detect if atomics are not necessary (i.e., accessed by only one thread).

**Messy Parameters:** When writing functions with parameters of type *mut T, we find that the parameter becomes long and confusing. For example, if we wanted to create a mutable reference to a mutable pointer, in other words, we are able to mutate both the pointer and the memory that the pointer points to, the function parameter might look something like:

```rust
fn foo(p: &mut *mut usize) {
    *p = Box::into_raw(Box::new(8usize)); //p points to 8
    unsafe { **p = 3 }; //p points to 3
}
```

This parameter is fairly confusing and it can be difficult to determine the purpose of the mut, *, and &. For a beginner with little experience with Rust's with Rust's pointer and reference types, many questions arise with this sort of function parameter. What is the difference between the two `mut`s? How does this type correspond to variable p? The overall complexity of the parameter creates confusion and can make it difficult to read and write Rust function parameters.

**Atomic Ordering:** When implementing concurrent code it is almost always necessary to use atomic variable types in order to ensure that concurrent modifications to data do not result in lost operations (e.g., `a += 1` vs `a.fetch_add(1)`). However, in Rust, almost all Atomic functions require the use of some type of `Ordering` as a parameter. When writing our concurrent hashmap, we did not use any other `Ordering` except `Ordering::SeqCst`, in part because the meaning of the different orderings were not clear to us. It is difficult to understand the implications of any particular Ordering, even after reading the documentation.

**Unnecessary Use of Lifetimes:** We noticed that often when returning Nodes from functions, if we implement our function incorrectly, as a beginner might, the compile will warn us about improper lifetimes and tell us to add them. This, however, is sometimes misleading as the correct solution will often not require explicit lifetimes to be added.

```rust
fn search(ptr: *mut Node) -> &Node {
    let mut node = &Node::new(None, None);
    let mut t = unsafe { &*self.head.load(Ordering::SeqCst) };
    node = t;
    return node;
}

fn search() -> &usize {
   &*Box::new(8usize)
}

fn main() {
   search();
}
```

https://play.rust-lang.org/?gist=bbafb9c6582f80152678c6b0ca25aa45&version=stable&mode=debug&edition=2015

The example above produces the error "error[E0597]: borrowed value does not live long enough" and that "note: borrowed value must be valid for the anonymous lifetime #1 defined on the method body". This implies that we might need a lifetime to fix this error, however one correct solution might not even need lifetimes, but rather changing the return type.

```
fn search(&self) -> *mut Node {
    let mut node: *mut Node = ptr::null_mut();
    let mut t = self.head.load(OSC);

    node = t;
    return node;
}

fn get_foos_bar(x: &mut Foo) -> &mut Bar {
  &mut x.bar
}
```

## Unsafe and Encapsulation

Unsafe is necessary for managing raw pointers, and for safely sharing data when sophisticated invariants that are outside the compiler's knowledge are maintained (e.g., the epoch-based memory reclamation strategy discussed above). However, encapsulation allows developers to limit the scope of that unsafety and present safe APIs instead to prevent a viral unsafe infection throughout the code. For example, in our concurrent hashmap implementation, we define the hashmap as a vector of linked list types that we implement. While our custom linked list requires unsafe code in order to move between nodes or to free deleted nodes, the hashmap type, which users interact with, only contains safe code.

Crossbeam is a Rust library that provides safe encapsulations for many common concurrency primitives like memory reclamation, lock-free stacks and queues, etc.. It isolates most of the unsafe elements of concurrent code to just that library, and allows authors of even sophisticated concurrent algorithms to eliminate most if not all uses of unsafe from their code. One example of this is deletion from the linked list buckets in our concurrent hashmap. In our initial from-scratch implementation, we had to design and implement an epoch-based memory reclamation scheme to safely deallocate nodes when they are deleted. Using Crossbeam, we were able to eliminate all unsafe blocks that we implemented from the linked list by using their safe atomic object APIs, which integrates automatically with an efficient epoch-based memory reclamation scheme. Crossbeam manages to present a safe API for memory reclamation by leveraging the Rust type system: developers are forced to "pin" an epoch to a stack variable before proceeding with any atomic operation in a function, which must outlive any of the values that are atomically read. This ensures that developers cannot even accidentally read a value without correctly incrementing the appropriate epoch, and that they correctly clean up the epoch after they have release all references read during that epoch. Crossbeam allowed us to write a safe remove method for our linked list using epoch-based reclamation without any bugs on the first try.

## Concluding Remarks

We implemented a lock-free concurrent hashmap available as a Rust crate that provides a significant speedup increasing the number of threads. More importantly, we highlighted the set of features used by the Rust language to highlight safety issues in code. We further examined where these features are helpful to developers, and where these features add conclusion and complexity in writing code. In the future, we would like to further research and potentially implement methods to address certain aspects that make writing concurrent code in Rust more difficult.

## Acknowledgements

We would like to thank Prof. Frans Kaashoek at MIT Parallel and Distributed Operating Systems Group for advising us over the course of the project and for directing us to focus on the Rust language due to its concurrency-aware design. We would also like to thank Prof. Srini Devadas and Dr. Slava Gerovitch for supporting us through the MIT PRIMES program.

## References

1. https://dl.acm.org/citation.cfm?id=3158154
2. https://www.microsoft.com/en-us/research/wp-content/uploads/2001/10/2001-disc.pdf
3. https://www.cse.wustl.edu/~angelee/archive/cse539/spr15/lectures/lists.pdf
4. http://lass.cs.umass.edu/~shenoy/courses/fall09/lectures/Lec15.pdf
5. https://hackernoon.com/eventual-vs-strong-consistency-in-distributed-databases-282fdad37cf7
6. https://docs.microsoft.com/en-us/windows/desktop/FileIO/synchronous-and-asynchronous-i-o
7. http://www.cecs.uci.edu/~papers/ipdps06/pdfs/1568974892-IPDPS-paper-1.pdf
8. https://doc.rust-lang.org/book/
9. https://doc.rust-lang.org/std/cell/index.html
10. https://doc.rust-lang.org/nomicon/concurrency.html
11. https://www.cl.cam.ac.uk/research/srg/netos/papers/2001-caslists.pdf